\documentclass[11pt]{article}
\usepackage{cite}
\usepackage{graphicx}
\usepackage{indentfirst}
\usepackage{amsmath}
\usepackage{amssymb}
\usepackage{amsthm}
\usepackage{amsfonts}

\usepackage{anysize}
\usepackage{fancyhdr}
\title{The Impact of Channel Feedback on Opportunistic Relay Selection for Hybrid-ARQ in Wireless Networks}
\author{Caleb K. Lo, Robert W. Heath, Jr. and Sriram Vishwanath \\ Wireless Networking and Communications Group \\ Department of Electrical and Computer Engineering \\ The University of Texas at Austin \\ 1 University Station C0803 \\ Austin, TX 78712-0240 \\ Phone: (512) 232-2014 \\ Fax: (512) 471-6512 \\ Email: \{clo, rheath, sriram\}@ece.utexas.edu}
\date{}

\begin{document}
\maketitle

\begin{abstract}
This paper presents a decentralized relay selection protocol for a dense wireless network and describes channel feedback strategies that improve its performance.  The proposed selection protocol supports hybrid automatic-repeat-request transmission where relays forward parity information to the destination in the event of a decoding error.  Channel feedback is employed for refining the relay selection process and for selecting an appropriate transmission mode in a proposed adaptive modulation transmission framework.  An approximation of the throughput of the proposed adaptive modulation strategy is presented, and the dependence of the throughput on system parameters such as the relay contention probability and the adaptive modulation switching point is illustrated via maximization of this approximation.  Simulations show that the throughput of the proposed selection strategy is comparable to that yielded by a centralized selection approach that relies on geographic information.
\end{abstract}

Keywords - Adaptive modulation, automatic repeat request, convolutional codes, relays.

\section{Introduction}
Relay-assisted communication is likely to play a major role in future-generation cellular systems and ad hoc networks, based on recent work for the IEEE 802.11s \cite{ESSMeshNetw} and IEEE 802.16j \cite{RelaTaskGrou} standards.  One form of relay-assisted communication stems from the system model in \cite{CovGam:CapaTheoRelaChan:Sep:79}, where direct transmission occurs between a source node and its corresponding destination node.  Given that a direct transmission has occurred, designated relay nodes can assist the destination in recovering the source message if the direct transmission fails.  Another form of relay-assisted communication is multihop relaying \cite{PabETAL:RelaBaseDeplConc:Sep:04, BoyETAL:MultDiveWireRela:Oct:04, DohETAL:CapaDistPhyLaye:Mar:06}, where transmission occurs between designated relays with the overall objective of forwarding the source message to its destination, and direct transmission between the source and the destination does not occur.  Yet another form of relay-assisted communication is cooperative diversity \cite{SenETAL:UserCoopDiveSyst:Nov:03, JanETAL:CodeCoopWireComm:Feb:04, LanWor:DistSpacTimeCode:Oct:03}, where multiple sources cooperate to send each source's message to a common destination.  In a cooperative diversity system, each transmitting node has its own message, while other forms of relay-assisted communication rely on designated relay nodes that do not have their own messages to transmit.  By employing designated relays in a cellular system or ad hoc network, critical performance improvements in terms of coverage extension, increased throughput, and higher system capacity can be realized \cite{RelaTaskGrou}.  Moreover, deploying designated relays in a wireless network reduces deployment and operation costs compared to the deployment of additional base stations \cite{RelaTaskGrou}.

Even though relay-assisted communication yields key performance and cost improvements, communication still occurs over inherently lossy wireless links.  Deep channel fades degrade the quality of the received packet, which leads to unacceptable decreases in throughput and reliability.  This resulting performance degradation can be mitigated by implementing hybrid-ARQ transmission strategies, including Chase combining and parity forwarding based on incremental redundancy \cite{ZhaVal:PracRelaNetwGene:Jan:05}.  For example, relays that are situated between the source and the destination can forward parity information to the destination if it detects uncorrectable packet errors, which leads to spectral efficiency gains \cite{ZhaVal:PracRelaNetwGene:Jan:05, RazYu:PariForwMultRela:Jul:06, PopCar:SpecEffiWireRela:Apr:07, RanWit:SpecEffiProtHalf:Feb:07}.  On average, the destination receives more reliable parity information from the relays than from the source, since the average channel between the source and the destination is worse than the average channel between each relay and the destination.  The increased reliability of the parity information improves the destination's ability to decode the source message, which decreases the number of retransmissions that are needed for successful decoding, resulting in reduced transmission delay and probability of buffer overflow at all receiving nodes.

In a relay-based wireless system, the source selects either a single relay or multiple relays to forward either its original message or parity information to the destination.  There has been significant prior work on single relay selection \cite{CheSerETAL:DistPoweAlloPara:Nov:05,LinErk:RelaSearAlgoCode:Nov:05,SadHanETAL:DistRelaAssiAlgo:Jun:06,ZhaAdvETAL:ImprAmplForwRela:Jul:06,BleKhiETAL:SimpCoopDiveMeth:Mar:06,LuoBluETAL:ApprCoopMultAnte:Sep:04,LiuSu:OptiSeleRelaProt:Nov:06,ZhaVal:PracRelaNetwGene:Jan:05} and multiple relay selection \cite{LanWor:DistSpacTimeCode:Oct:03, DohETAL:CapaDistPhyLaye:Mar:06, JinHas:DistSpacTimeCodi:Dec:06}.  In this paper we focus on single-relay selection for several reasons.  For example, it was shown in \cite{BleKhiETAL:SimpCoopDiveMeth:Mar:06} that by selecting the relay with the best instantaneous channel gain to the destination, a diversity gain on the order of the number of relays in the network could be realized.  This important result reduces the need for implementing multiple-relay strategies such as distributed space-time coding \cite{JinHas:DistSpacTimeCodi:Dec:06} and distributed beamforming.  Even though diversity gain is defined in the high-SNR regime, it is a useful benchmark for system performance in the low and medium-SNR regimes.  Distributed space-time coding suffers from drawbacks such as the difficulty of synchronizing transmissions from disparate nodes and designing good codes that are easy to implement.  Distributed beamforming is difficult to implement in practice since the oscillators in distinct nodes are not necessarily synchronized and are subject to phase noise.  Note that a simple multiple-relay strategy that involves the decoding relays forwarding their parity information in orthogonal time slots also suffers from the difficulty of transmission synchronization.  As the number of decoding relays increases, more time slots must be dedicated to parity forwarding, which increases transmission delay.

In terms of single-relay selection, \cite{ZhaVal:PracRelaNetwGene:Jan:05} is the most closely related work to this paper.  The proposed selection strategy in \cite{ZhaVal:PracRelaNetwGene:Jan:05} relies on either Global Positioning System (GPS) information or relays overhearing a sufficient number of ACK messages from the destination to select the closest decoding relay to the destination to forward parity information.  This selection method optimizes the average SNR at the destination, but practical implementation is difficult since each relay is required to know its average SNR to the destination and the average SNR to the destination for all other relays.  This difficulty is only exacerbated as the number of nodes increases.  Thus, a more decentralized method for relay selection is preferable.

In this paper, we propose a decentralized relay selection approach that involves random access-based feedback to the source \cite{LoETAL:HybrARQMultNetw:Apr:07, LoETAL:OppoRelaSeleLimi:Apr:07}.  The uniqueness of this paper is that the relay selection strategy is based on $\textit{opportunistic}$ $\textit{feedback}$, which is applied to user scheduling in a downlink multiuser system in \cite{TanHea:OppoFeedDownMult:Oct:05}.  Our strategy is based on the system model in \cite{CovGam:CapaTheoRelaChan:Sep:79}, where transmission occurs over the direct link between the source and the destination.  In our approach, the source uses ACK messages from the relays to select a relay to forward parity information if the destination cannot recover the source message based on the initial direct-link transmission.  Each ACK message is an identification (ID) number that is unique to each relay.  The number of ACK messages from the relays are controlled by factors including the contention probability for each relay.  The contention probability is the probability that a decoding relay with a channel gain to the destination above a given threshold $\eta_{opp}$ sends an ACK message to the source.  We illustrate the impact on system performance of varying key parameters such as $\eta_{opp}$ and the contention probability.

We also further refine the relay selection process by appending a flag bit to each ACK message \cite{LoETAL:OppoRelaSeleLimi:Apr:07}, where the value of this flag bit is found by partitioning the set of relay channel gains to the destination according to a given threshold $\beta_{opp} > \eta_{opp}$ and determining to which partition each decoding relay belongs.  The flag bit determines a subset of the decoding relays such that the selection process is biased in favor of the relays in this subset.  Refining the relay selection process closes the performance gap between our selection strategy and centralized strategies that select the decoding relay with the best channel gain to the destination \cite{BleKhiETAL:SimpCoopDiveMeth:Mar:06}.

Our proposed strategy as outlined in \cite{LoETAL:HybrARQMultNetw:Apr:07, LoETAL:OppoRelaSeleLimi:Apr:07} relies on the use of rate-compatible punctured convolutional (RCPC) coding \cite{Hag:RateCompPuncConv:Apr:88}, where the source transmits using a high-rate code and then the relays contend to forward parity information so that the destination can decode the source message using successively lower-rate codes.  As noted above, channel feedback in the form of a flag bit in each ACK message is used to refine the relay selection process.  We also employ relay channel feedback in an adaptive modulation transmission framework \cite{LiuETAL:CrosLayeCombAdap:Sep:04, CatETAL:AdapModuMIMOCodi:Jun:02, GolChu:AdapCodeModuFadi:May:98}.  Whenever a node is about to transmit to the destination, it measures its channel gain to the destination and uses this value to determine an appropriate transmission mode.  We demonstrate that this adaptive modulation and coding (AMC) approach yields throughput gains over a strategy that uses a single transmission mode.  We also derive an approximation of the throughput of our AMC approach.  By maximizing the approximation over parameters such as the AMC switching point and the relay contention probability, we obtain optimal values for these parameters that can be used by system designers to maximize the throughput of the proposed strategy.  Even though our proposed selection strategy relies on instantaneous channel state information (CSI), we derive throughput approximations to evaluate the long-term performance of our proposed strategy when employing AMC.

Our proposed strategy with RCPC coding significantly outperforms a point-to-point hybrid-ARQ strategy where the source always forwards parity information to the destination.  Also, our proposed strategy with RCPC coding yields throughputs that are comparable to those given by the HARBINGER strategy in \cite{ZhaVal:PracRelaNetwGene:Jan:05}, which demonstrates that decentralized selection strategies with reduced signaling overhead can still offer good performance.  One reason for this result is that HARBINGER does not consider instantaneous CSI, since relay selection is based on proximity to the destination.  On the other hand, the ACK messages from the relays in our proposed strategy are controlled by the instantaneous channel gains from the relays to the destination.  In fact, by performing a difficult joint optimization over several key system parameters, it is possible for our proposed strategy to outperform HARBINGER and approach the performance of the instantaneous CSI-based method in \cite{BleKhiETAL:SimpCoopDiveMeth:Mar:06}.  Also, as noted above, using a limited amount of channel feedback closes the performance gap between our proposed strategy and the method in \cite{BleKhiETAL:SimpCoopDiveMeth:Mar:06}.  Even though our strategy cannot outperform the optimal method in \cite{BleKhiETAL:SimpCoopDiveMeth:Mar:06}, it can be readily integrated into a practical hybrid-ARQ wireless system.  For example, punctured coding techniques are being employed in the IEEE 802.16e standard \cite{WireMANWorkGrp}.  On the other hand, practical code designs are not discussed in \cite{BleKhiETAL:SimpCoopDiveMeth:Mar:06}.


This paper is organized as follows.  In Section II we describe the system model and our proposed relay selection protocol.  In Section III, we modify the proposed protocol to support adaptive modulation and coding.  An approximation of the throughput of the proposed adaptive modulation approach is presented in Section IV.  Simulation results are presented in Section V.  We conclude the paper in Section VI.

\section{System Model}\label{sysmod}
We use boldface notation for vectors.  SNR represents the signal-to-noise ratio.  $\|\mathcal{A}\|$ denotes the cardinality of a set $\mathcal{A}$.  $|z|^2$, $z^{*}$, $\Re(z)$ and $\Im(z)$ denote the absolute square, complex conjugate, real part and imaginary part, respectively, of a complex number $z$.  For a real number $n$, $\lceil n \rceil$ denotes the smallest integer $n_0$ such that $n_0 \geq n$.  $\mathbb{E}(X)$ represents the mathematical expectation of the random variable $X$.

\begin{figure}[h]
\begin{center}
\includegraphics[width=3.0in]{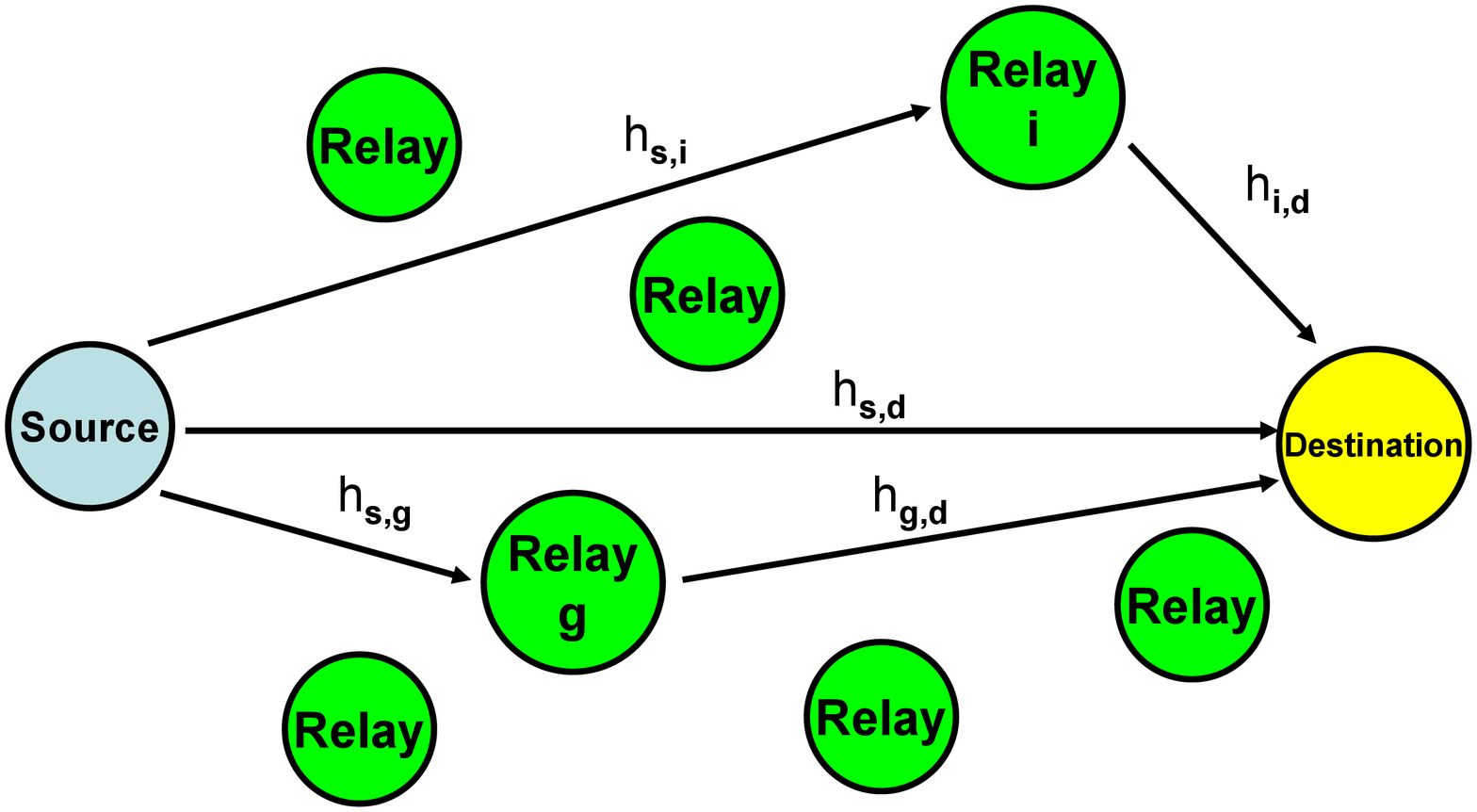}
\end{center}
\caption{Relay network.}
\label{system-model}
\end{figure}

The system of interest is shown in Fig. \ref{system-model}.  There are $K_r$ relays in the region between the source and the destination.  Each relay is equipped with a single antenna.

In the proposed protocol, data transmission occurs over a set of time slots $\{t_1,...,t_m\}$, which are of variable duration.  The duration of each time slot depends on the particular modulation and code rate employed by the transmitting node along with the amount of data that is to be transmitted during that time slot.  The source initially has a k-bit message $w$ that is encoded as an n-bit codeword $\textbf{x}(w)$.  Before the first time slot $t_1$, the source and destination perform request-to-send/clear-to-send (RTS/CTS) handshaking to achieve synchronization.  RTS/CTS handshaking also synchronizes all $K_r$ relays with the impending transmission between the source and the destination, where we assume that all relays lie within the transmission range of both the source and the destination.  The relays will overhear both the RTS and CTS messages and prepare to receive the source's transmission in $t_1$.  Note that the classic hidden terminal problem arises if any relays are within the interference range of either the source or the destination \cite{KumETAL:CommNetwAnalAppr:04}.

Then, at the start of $t_1$, the source transmits a subset $\textbf{x}_1(w)$ of the bits in $\textbf{x}(w)$.  Let $h_{s,i,j}$ be the Rayleigh fading coefficient for the channel between the source and node $i$ during $t_j$, and let $\textbf{n}_{i,j}$ be additive white Gaussian noise with variance $N_0$ at node $i$ during $t_j$.  The destination observes
\begin{equation}
\textbf{y}_{d,1} = h_{s,d,1}\textbf{x}_1(w) + \textbf{n}_{d,1}
\end{equation}
while relay $i \in \{1,2,...,K_r\}$ observes
\begin{equation}
\textbf{y}_{i,1} = h_{s,i,1}\textbf{x}_1(w) + \textbf{n}_{i,1}.
\end{equation}
We use the Rayleigh fading distribution for our throughput derivations in Section \ref{thr-app} and in our simulations in Section \ref{sim-res}.

After the destination observes $\textbf{y}_{d,1}$, it attempts to recover $w$.  If the destination successfully recovers $w$, it broadcasts an acknowledgment (ACK) message to all of the relays and the source.  On the other hand, if the destination cannot recover $w$, it broadcasts a negative acknowledgment (NACK) message to all of the relays and the source, and the source attempts to select one of the relays to forward additional parity information that will assist the destination in recovering $w$.  In Section \ref{sel} we describe our relay selection protocol.

\subsection{Key Assumptions}
We assume that the reciprocity principle holds, so each node that is currently in a receiving mode learns its fading coefficient with the transmitting node via training data at the beginning of each transmission.  We also assume that all relays lie within the transmission range of both the source and the destination, which facilitates RTS/CTS handshaking that mitigates the hidden terminal problem and synchronizes the relays in our proposed selection strategy.  In addition, we assume that each relay operates in a half-duplex mode, so none of the relays can simultaneously transmit and receive.

We make a block fading assumption here, i.e. that all fading coefficients are constant over a time slot and vary independently from slot to slot.  This is a reasonable assumption assuming that each time slot is much less than the channel coherence time.  We also assume that the fading coefficients and additive noise realizations are independent across the nodes; these are reasonable assumptions assuming that the separation between any two nodes in our network is greater than the channel coherence distance.

\subsection{Relay Selection}\label{sel}
\begin{figure}[tb]
\begin{center}
\includegraphics[width=3.0in]{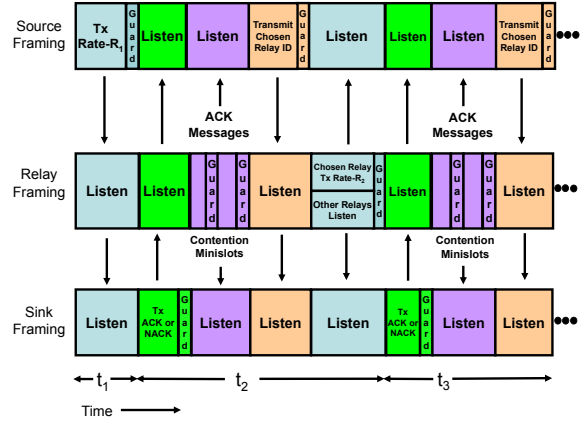}
\end{center}
\caption{Framing structure for proposed selection strategy.}
\label{framing-structure}
\end{figure}

We modify the opportunistic feedback approach in \cite{TanHea:OppoFeedDownMult:Oct:05} to select one of the relays for transmission in $t_2$.  An example of the medium access control (MAC) layer framing structure for our protocol is shown in Fig. \ref{framing-structure}.  After transmission from the source in $t_1$, we assume in Fig. \ref{framing-structure} that the destination broadcasts a NACK message to start the contention process; the same is true for $t_2$.

Let $\mathcal{R}_{sel}$ denote the set of relays that can participate in the relay contention process, where relay $i \in \mathcal{R}_{sel}$ has both recovered $w$ and has a channel gain to the destination $|h_{i,d,2}|^2$ that is above a threshold $\eta_{opp}$.  In Section \ref{perf-impact} we demonstrate the performance impact of varying the parameter $\eta_{opp}$.  Each relay $i$ will determine $|h_{i,d,2}|^2$ by listening to the destination's NACK message at the beginning of $t_2$.  All relays in $\mathcal{R}_{sel}$ are allocated the same $K$ minislots for feedback to the source.

Note that we have assumed a block fading channel model and that each decoding relay determines its channel gain to the destination by listening to the destination's NACK message.  This implies that in our proposed framing structure, each time slot must begin with either an ACK or a NACK from the destination for the transmission during the preceding time slot.  This is illustrated in Fig. \ref{framing-structure}.  If we were to adopt the framing structure of conventional MAC protocols that assume that a time slot concludes with an ACK or NACK from the destination \cite{ESSMeshNetw}, the block fading assumption would imply that a selected relay might have a poor channel gain to the destination during its transmission.  Thus, if we were to adopt a more conventional framing structure, we would need to assume a fading channel model with time correlation.

After the destination broadcasts a NACK message, the contention period begins.  During contention minislot $b$, each relay $i \in \mathcal{R}_{sel}$ will send a ACK message to the source with probability $p_i$, where this ACK message consists of an ID number that has been assigned to relay $i$.  Successful contention occurs during minislot $b$ if exactly one relay $i \in \mathcal{R}_{sel}$ sends a ACK message to the source, and relay $i$ is declared to be the $\textit{winner}$ for minislot $b$.  If relays $s, t \in \mathcal{R}_{sel}$ send ACK messages during minislot $b$ and $s \neq t$, a collision occurs and the source discards all received ACK messages.  After minislot $K$ has been completed, the source determines if at least one $\textit{winning}$ relay exists.  If so, the source randomly selects one of the $\textit{winning}$ relays $i_{t}$ to transmit during $t_2$.  If there are no $\textit{winning}$ relays, the source will transmit during $t_2$.  The source then broadcasts the ID number of the node that will transmit during $t_2$.

Note that a guard interval is included at the end of each contention minislot.  Each guard interval is equal to the propagation delay of the network in Fig. \ref{system-model}.  The purpose of the guard interval is to prevent ACK messages that are sent during a minislot from colliding with ACK messages that are sent during the next minislot.  Guard intervals are also included after the following events: 1) a selected node transmits, 2) the destination transmits an ACK or a NACK, and 3) the source transmits the ID number of the selected node.  These guard intervals are included to facilitate network synchronization.

During $t_2$, relay $i_{t}$ (or the source) transmits a subset $\textbf{x}_2(w)$ of the bits in $\textbf{x}(w)$.  The destination should not discard $\textbf{y}_{d,1}$ after $t_1$, but it should combine $\textbf{y}_{d,1}$ with
\begin{equation}
\textbf{y}_{d,2} = h_{i_{t},d,2}\textbf{x}_2(w) + \textbf{n}_{d,2}
\end{equation}
and attempt to recover $w$ from the combined output.  We describe two methods of combining $\textbf{y}_{d,1}$ with $\textbf{y}_{d,2}$ in Section \ref{sel-rcpc} and Section \ref{sel-amc}.  If decoding at the destination is unsuccessful, the destination broadcasts another NACK message to all of the relays and the source, and then the relay contention process is repeated to select another relay to transmit during $t_3$.  The retransmission and contention processes repeat until the destination either successfully recovers $w$ or fails to recover $w$ after $t_m$ has elapsed.

We remark that our proposed strategy is distinct from the ``instantaneous-relaying'' and ``random-relaying'' approaches that are proposed in \cite[Section III.B]{ZhaVal:PracRelaNetwGene:Jan:05}.  The ``instantaneous-relaying'' strategy always chooses the relay with the best instantaneous channel gain to the destination, which is the same strategy that is employed in \cite{BleKhiETAL:SimpCoopDiveMeth:Mar:06}.  In contrast, our strategy chooses a relay with a good instantaneous channel gain to the destination, but it may not necessarily choose the relay with the best instantaneous channel depending on the outcome of the contention process.  As for the ``random-relaying'' strategy, each relay that has decoded the source message will transmit in the next time slot with a certain probability.  On the other hand, our strategy ensures that exactly one node will be selected to transmit in the next time slot based on the outcome of the contention process, and only those decoding relays that have good channel gains to the destination will even be able to participate in the contention process.

\subsection{RCPC Signaling}\label{sel-rcpc}
We adopt the ARQ/FEC protocol in \cite[Section 5.A]{Hag:RateCompPuncConv:Apr:88}, so the source chooses code rates $\{R_1,R_2,...,R_m\}$ from a RCPC family, and $R_1 > R_2 > \cdots > R_m$.  The rate-$R_m$ code is the mother code of the RCPC family.

The rate-$R_m$ code is used to encode $w$ as a codeword $\textbf{x}(w)$.  During $t_1$, the source transmits a subset $\textbf{x}_1(w)$ of the bits in $\textbf{x}(w)$ such that $\textbf{x}_1(w)$ forms a codeword from the rate-$R_1$ code.  Then the destination attempts to decode $\textbf{y}_{d,1}$ based on the rate-$R_1$ code.

If unsuccessful decoding occurs, during $t_2$, the chosen relay (or the source) transmits a subset $\textbf{x}_2(w)$ of the bits in $\textbf{x}(w)$ such that $\textbf{x}_1(w) \cup \textbf{x}_2(w)$ forms a codeword from the rate-$R_2$ code.  Then the destination attempts to decode $\textbf{y}_{d,1} \cup \textbf{y}_{d,2}$ based on the rate-$R_2$ code.  This continues until either $w$ is recovered at the destination or $\textbf{x}(w)$ is transmitted without successful recovery of $w$.

Let $M$ be the memory of the mother code and let $l_{AV}$ be the average number of additionally transmitted bits per $P$ information bits, where $P$ is the puncturing period of the RCPC family.  To compute the dimensionless effective code rate of this strategy, we use \cite[equation (16)]{Hag:RateCompPuncConv:Apr:88}
\begin{equation}\label{hagenauer-throughput}
R_{avg} = \bigg(\frac{k}{n+M}\bigg) \bigg(\frac{P}{P+l_{AV}}\bigg).
\end{equation}
We refer to $R_{avg}$ as the throughput of this strategy in the rest of this paper.

\subsection{Channel Feedback for Refining Relay Selection}\label{chan-fbk-rel-sel}
In the proposed relay selection strategy, each relay's ACK message consists of an ID number that has been assigned to it.  We refer to our selection approach as an $\textit{ID}$ strategy.  Now it is possible to modify the $\textit{ID}$ strategy by appending a flag bit to the ACK message from relay $i$, where the flag bit is set to `1' only if $|h_{i,d,j}|^2 > \beta_{opp}$ for $\beta_{opp} > \eta_{opp}$.  Again, successful contention occurs during minislot $b$ if exactly one relay $i \in \mathcal{R}_{sel}$ sends a ACK message.  We refer to this approach as an $\textit{ID-CSI-1}$ strategy.

Note that the $\textit{ID-CSI-1}$ approach is a 1-bit channel feedback strategy, since the channel space for each contending relay is partitioned into two sets according to $\beta_{opp}$ and each relay sets its flag bit according to the set that contains its channel gain to the destination.  This approach can be generalized to an $N$-bit channel feedback strategy, where $N$ is an arbitrary positive integer.  In this case, the channel space for each contending relay would be partitioned into $2^N$ sets according to a set of thresholds $\beta_{opp,1} > \beta_{opp,2} > \cdots > \beta_{opp,2^N-1} > \eta_{opp}$.  Each contending relay would set its $N$ flag bits according to the set that contains its channel gain to the destination.  In the rest of the paper, we focus on the 1-bit channel feedback strategy.

After minislot $K$, if either all of the winners sent a flag bit of '0', all of the winners sent a flag bit of '1', or there are no winners, the $\textit{ID-CSI-1}$ strategy reduces to the $\textit{ID}$ strategy.  Otherwise, the source will randomly select one of the winners $i_{t}$ that sent a flag bit of '1' with probability $q > 0.5$.  One of the winners $i_{t}$ that sent a flag bit of '0' is randomly selected with probability $1-q$.  Thus, the $\textit{ID-CSI-1}$ strategy refines the $\textit{ID}$ strategy by further biasing the selection process in favor of the relays with very good channel gains to the destination.

By choosing a value of $\beta_{opp}$ we determine a subset of the decoding relays that have a better chance of being selected.  In Section \ref{perf-impact} we present simulation results that show how the throughput of the $\textit{ID-CSI-1}$ strategy varies as a function of $\beta_{opp}$.

\section{Relay Selection with Adaptive Modulation}\label{sel-amc}
The transmission strategy in Section \ref{sel-rcpc} relied on the use of RCPC coding.  Transmissions during successive time slots consisted of parity bits encoded in the same fixed modulator that allowed the destination to decode the source message using successively lower-rate codes.

Note that RCPC coding has the advantage of allowing for fine rate control, since by sending a limited number of parity bits in successive time slots, the monotonic decrease in code-rate is mitigated.  Decoding based on puncturing rules is difficult to implement, though, and additional memory is required at each node to store the puncturing tables for the RCPC family.  Another transmission strategy that the nodes in the relay network can use in conjunction with hybrid-ARQ is adaptive modulation and coding (AMC).  It is relatively straightforward to implement AMC at each node in the relay network since AMC allows for simple decoding strategies such as Chase combining \cite{Cha:CodeCombMaxiLike:May:85}.  We note that transmission involving both AMC and code puncturing is possible as evidenced by the IEEE 802.16e standard \cite{WireMANWorkGrp}, though we do not consider a combination of both strategies in this paper.

We propose the following transmission strategy that is based on AMC.  Assume that each transmitting node can choose from a total of $N$ transmission modes.  In Sections \ref{chase-comb-sm}, \ref{chase-comb-tm} and \ref{thr-app} we describe and analyze the performance of our proposed strategy for the special cases of $N = 1$ and $N = 2$ transmission modes.  Each transmission mode $i$ consists of a particular modulation/code-rate pair.  Define a set of channel power thresholds $\{\gamma_1,\gamma_2,\ldots,\gamma_{N-1}\}$, where $0 < \gamma_1 < \gamma_2 < \cdots < \gamma_{N-1}$.

During each time slot $t_a$, the selected node $i_t$ measures its channel gain to the destination $|h_{i_t,d,a}|^2$.  If $\gamma_{j-1} < |h_{i_t,d,a}|^2 \leq \gamma_j$, then this node will encode and modulate the source message using transmission mode $j$.  The encoded and modulated message is then sent to the destination.  Note that the dimensionless effective code rate is a random variable since the number of time slots and the transmission mode that is used in each time slot depends on the channel gain $|h_{i_t,d,a}|^2$ in each time slot.  We refer to the dimensionless effective code rate as the throughput in the rest of this paper.  The objective here, as is generally the case with adaptive modulation, is to maximize the expected throughput.

The relay selection protocol as described in Section \ref{sel} is still used here; it is just the transmission strategy that the selected relay $i_t$ (or the source) uses that is different, and each receiving node uses Chase combining on the received packets to decode the source message.  Chase combining is a soft-decision maximal-ratio combining (MRC) decoding strategy.

\subsection{Chase Combining with Single Transmission Mode}\label{chase-comb-sm}
During a set of time slots $\{t_1,t_2,\ldots,t_m\}$ where the destination attempts to recover $w$, the transmitting node in each time slot can employ any of the $N$ transmission modes.  Initially we assume that only one mode, say mode $J$, is used during $\{t_1,t_2,\ldots,t_m\}$.  Later, we will present an example where this assumption is relaxed and we describe how our Chase combining approach is modified.  Consider time slot $t_j$ where node $i$ is in a receiving mode.

After the received packet is de-interleaved, it is combined with the previous received packets using MRC.  The MRC estimate for each transmitted bit is then used to compute a likelihood ratio for that bit.  Then, the likelihood ratio for each bit is quantized and passed to a soft-decision Viterbi decoder to recover the source message.


As noted in \cite{Cha:CodeCombMaxiLike:May:85}, node $i$ should discard any received packets that have an error rate that is approximately 1/2.  Since the error rate cannot be measured directly from the received packet, the following equivalent approach is used: if the received SNR at node $i$ is less than $\phi$, node $i$ discards the received packet.  It is conceivable that node $i$ could keep discarding packets if the received SNR stays below the threshold $\phi$, so we set a limit on the number of time slots that are allowed for a particular source message before the destination stops attempting to decode it.  Similarly, we set a limit on the number of decoding attempts that the destination can make for a particular source message.

We assume that if the received SNR at node $i$ is greater than $\lambda_J$, node $i$ is able to decode the source message.  Now, since MRC-based Chase combining is being used at each receiving node, the SNR of the combined packet at node $i$ improves as the number of retransmissions increases.  The objective of Chase combining is to increase the received SNR of the combined packet at node $i$ until it exceeds $\lambda_J$.

\subsection{Chase Combining Example with Two Transmission Modes}\label{chase-comb-tm}
To illustrate our adaptive modulation approach, consider a link adaptation algorithm with $N = 2$ transmission modes.  Let Mode 1 be a combination of binary phase-shift-keying (BPSK) modulation with a rate-1/3 convolutional code and let Mode 2 be a combination of quadrature phase-shift-keying (QPSK) modulation with a rate-2/3 convolutional code.  When transmission occurs over bad channel conditions, Mode 1 is employed to yield good error performance, since it combines a low-rate code with a low symbol rate.  On the other hand, when transmission occurs over good channel conditions, Mode 2 is employed to yield good throughput performance, since it combines a high-rate code with a higher symbol rate than that used in Mode 1.

To facilitate the computation of MRC estimates at receiving node $i$ when both Mode 1 and Mode 2 have been employed, let the rate-1/3 code for Mode 1 be a systematic convolutional code and the rate-2/3 code be a punctured version of this rate-1/3 code.

During time slot $t_a$, the transmitting node $i_t$ measures its channel gain to the destination $|h_{i_t,d,a}|^2$.  If $|h_{i_t,d,a}|^2 \geq \gamma_{swp}$, then this node will encode and modulate the source message $w$ using Mode 2.  On the other hand, if $|h_{i_t,d,a}|^2 < \gamma_{swp}$ then this node will encode and modulate the source message $w$ using Mode 1.  The encoded and modulated message is then sent to the destination.

\subsubsection{Employed Transmission Modes: Only Mode 1}\label{mode-1}
If only Mode 1 has been used thus far, the objective of Chase combining is to increase the received SNR of the combined packet until it exceeds $\lambda_1$, where $\lambda_1$ is the minimum SNR decoding threshold for Mode 1.  Decoding the source message is straightforward and follows the general guidelines in Section \ref{chase-comb-sm}.

\subsubsection{Employed Transmission Modes: Only Mode 2}\label{mode-2}
If only Mode 2 has been used thus far, the objective of Chase combining is to increase the received SNR of the combined packet until it exceeds $\lambda_2$, where $\lambda_2$ is the minimum SNR decoding threshold for Mode 2.  Note that $\lambda_2 > \lambda_1$.  Decoding the source message follows the general guidelines in Section \ref{chase-comb-sm}, except that each receiving node $i$ unpunctures the set of quantized soft-decision bit estimates to decode based on the mother rate-1/3 code to yield better error performance.  Forming the MRC estimate for each transmitted bit from the rate-1/3 code is not as straightforward as in Section \ref{mode-1}, so each receiving node $i$ employs the fact that the rate-2/3 code for Mode 2 is punctured from the rate-1/3 code to form the MRC estimates.

\subsubsection{Employed Transmission Modes: Modes 1 And 2}
If both Mode 1 and Mode 2 have been used thus far, the objective of Chase combining is to increase the received SNR of the combined packet until it exceeds $\lambda_1$.  Decoding the source message follows the general guidelines in Section \ref{mode-2}, where decoding is based on the mother rate-1/3 code.

\section{Throughput Approximation}\label{thr-app}
Given the relay selection strategy with adaptive modulation and Chase combining as presented in Section \ref{sel-amc}, we now present an approximation of the throughput of this strategy.  For ease of presentation, we assume that the destination must decode the source message within two time slots, or the source message will be discarded.  This is analogous to truncated ARQ, where transmission delay and buffer overflow are reduced by specifying a maximum number of retransmissions.  We also consider the special case of $N = 2$ transmission modes and employ the specific modulation/code-rate pairs in Section \ref{chase-comb-tm}.  Let $p_{i,j}$ denote the probability that the destination decodes the source message at the end of time slot $t_j$ given that transmission mode $i$ was used during time slot $t_j$.  Also, let $q_{a,i}$ denote the probability that relay $a$ is selected by the source after a time slot where transmission mode $i$ was used.  Let $q_{0,i} = 1-\sum_{a=1}^{K_r}q_{a,i}$ denote the probability that no relays are chosen by the source after a time slot where transmission mode $i$ was used.

Let the average received power at receiving node $j$ after a transmission from node $i$ be $|G_{i,j}|^2 = \mathcal{E}\cdot\mathbb{E}(|h_{i,j,a}|^2)$, where $\mathcal{E}$ is the transmit energy.  Recall that $\gamma_{swp}$ is the AMC switching point.  Let $\alpha$ and $\beta$ be the minimum SNR decoding thresholds for Modes 1 and 2, respectively.  Now
\begin{eqnarray}
p_{1,1} & = & \int_{\alpha}^{\gamma_{swp}}\frac{1}{|G_{s,d}|^2}e^{-\chi/|G_{s,d}|^2}d\chi \nonumber \\
& = & e^{-\alpha/|G_{s,d}|^2}-e^{-\gamma_{swp}/|G_{s,d}|^2},
\end{eqnarray}
\begin{eqnarray}
p_{2,1} & = & \int_{\beta}^{\infty}\frac{1}{|G_{s,d}|^2}e^{-\chi/|G_{s,d}|^2}d\chi \nonumber \\
& = & e^{-\beta/|G_{s,d}|^2},
\end{eqnarray}
\begin{eqnarray}
p_{1,2} & = & \int_{\phi}^{\gamma_{swp}}\Bigg(\sum_{a=1}^{K_r}\frac{1}{|G_{a,d}|^2}e^{-\chi/|G_{a,d}|^2}q_{a,1}+\frac{1}{|G_{s,d}|^2}e^{-\chi/|G_{s,d}|^2}q_{0,1}\Bigg)d\chi \nonumber \\
& = & \sum_{a=1}^{K_r}q_{a,1}\Bigg(e^{-\phi/|G_{a,d}|^2}-e^{-\gamma_{swp}/|G_{a,d}|^2}\Bigg)+q_{0,1}\Bigg(e^{-\phi/|G_{s,d}|^2}-e^{-\gamma_{swp}/|G_{s,d}|^2}\Bigg)
\end{eqnarray}
and finally
\begin{eqnarray}
p_{2,2} & = & \int_{\gamma_{swp}}^{\infty}\Bigg(\sum_{a=1}^{K_r}\frac{1}{|G_{a,d}|^2}e^{-\chi/|G_{a,d}|^2}q_{a,2}+\frac{1}{|G_{s,d}|^2}e^{-\chi/|G_{s,d}|^2}q_{0,2}\Bigg)d\chi \nonumber \\
& = & \sum_{a=1}^{K_r}q_{a,2}e^{-\gamma_{swp}/|G_{a,d}|^2}+q_{0,2}e^{-\gamma_{swp}/|G_{s,d}|^2}.
\end{eqnarray}

In the following examples we use code concatenation with an outer code with rate $R < 1$, and we zero-pad the codeword from the concatenated code with $M$ bits, where $M$ is the memory of the inner convolutional code, to bring the Viterbi decoder back to its all-zero state. Let $f$ denote the dimensionless effective rate of the outer code; the throughput $R_{amc}$ is approximated by
\begin{equation}\label{r-app-amc}
R_{app,amc} = \frac{f}{3}p_{1,1}+\frac{2f}{3}p_{2,1}+\frac{f}{3}(1-p_{2,1})p_{2,2}+\frac{f}{6}(1-p_{1,1})p_{1,2}+\frac{2f}{9}(1-p_{1,1})p_{2,2}+\frac{2f}{9}(1-p_{2,1})p_{1,2}.
\end{equation}

Now let $q_{a,i,j}$ denote the probability that relay $a$ wins $j$ out of $K$ minislots and is selected by the source, given that transmission mode $i$ was just used.  Thus
\begin{equation}
q_{a,i} = \sum_{j=1}^{K}q_{a,i,j}.
\end{equation}
Let $\mathcal{S} = \{S_1,S_2,\ldots,S_{\binom{K}{j}}\}$ denote the set of all subsets of $\mathcal{K} = \{1,2,\ldots,K\}$ that have cardinality $j$.  Consider $b \in \{0,1,\ldots,K-j\}$.  Let $\mathcal{B} = \{B_1,B_2,\ldots,B_{\binom{K-j}{b}}\}$ denote the set of all subsets of $\mathcal{K}\setminus S_m$ that have cardinality $b$, where $S_m \in \mathcal{S}$.  Let $u_{a,k,i}$ denote the probability that relay $a$ wins minislot $k \in \mathcal{K}$ given that transmission mode $i$ was just used.

Let $s_d \in S_c$ for $S_c \in \mathcal{S}$, $b_z \in B_v \in \mathcal{B}$ and $z_{\sigma} \in \mathcal{K}\setminus(S_c\bigcup B_v)$, so
\begin{equation}
q_{a,i,j} = \sum_{c=1}^{\binom{K}{j}}\Bigg(\Bigg[\prod_{d=1}^{j}u_{a,s_d,i}\Bigg]\Bigg[\sum_{m=0}^{K-j}\frac{j}{K-m}\Bigg\{\sum_{v=1}^{\binom{K-j}{m}}\Bigg(\prod_{z=1}^{m}\Bigg(1-\sum_{\kappa=1}^{K_r}u_{\kappa,b_z,i}\Bigg)\prod_{\sigma=1}^{K-j-m}\Bigg(\sum_{\psi\neq j}u_{\psi,z_{\sigma},i}\Bigg)\Bigg)\Bigg\}\Bigg]\Bigg).
\end{equation}

Let $\rho_{a,i}$ denote the probability that relay $a$ decodes the source message given that transmission mode $i$ was just used.  Now we note that $u_{a,k,i}$ is identical for all minislots $k \in \mathcal{K}$.  Thus, we can drop the subscript $k$
\begin{equation}
u_{a,i} = p_a\rho_{a,i}\sum_{R\subseteq (\mathcal{K}\setminus\{a\})}\Bigg(\prod_{c\in R}(1-p_c)\rho_{c,i}\prod_{d\in (\mathcal{K}\setminus(\mathcal{R}\cup\{a\}))}(1-\rho_{d,i})\Bigg).
\end{equation}
We note that
\begin{eqnarray}
\rho_{a,1} & = & \int_{\alpha}^{\infty}\frac{1}{|G_{s,a}|^2}e^{-\chi/|G_{s,a}|^2}d\chi \nonumber \\
& = & e^{-\alpha/|G_{s,a}|^2}
\end{eqnarray}
and
\begin{eqnarray}
\rho_{a,2} & = & \int_{\beta}^{\infty}\frac{1}{|G_{s,a}|^2}e^{-\chi/|G_{s,a}|^2}d\chi \nonumber \\
& = & e^{-\beta/|G_{s,a}|^2}.
\end{eqnarray}

To illustrate the throughput gains yielded by the AMC approach, we define another transmission strategy in our relay network that does not use AMC.  For this single-mode approach, during each time slot, the transmitting node $i_t$ encodes and modulates the source message $w$ using the same code/modulation pair.  The encoded and modulated message is then sent to the destination.

Forming the MRC estimates for each transmitted bit is fairly straightforward in this case, since only a single transmission mode is employed.  For example, consider a transmission mode that consists of a rate-1/2 code with generator polynomial (133 171) using octal notation and constraint length 7.  Each transmitting node uses quadrature-amplitude-modulated (QAM) signaling; in particular each transmitting node uses a 16-QAM constellation.  The decoding procedure at each receiving node $i$ follows the general guidelines in Section \ref{chase-comb-sm}.


As in the AMC approach, we assume that if the received SNR at node $i$ exceeds a minimum value $\gamma$, the receiving node $i$ is able to decode the source message.  The objective of Chase combining is to repeatedly combine the received packets until the received SNR of the combined signal at node $i$ is at least $\gamma$.

To approximate the throughput of this single-mode approach, let $\tau_j$ denote the probability that the destination decodes the source message at the end of time slot $t_j$.  Let $q_a$ denote the probability that relay $a$ is selected by the source during a time slot, and let $q_0 = 1-\sum_{a=1}^{K_r}q_a$ denote the probability that no relays are chosen by the source during a time slot.

Now
\begin{eqnarray}
\tau_1 & = & \int_{\gamma}^{\infty}\frac{1}{|G_{s,d}|^2}e^{-\chi/|G_{s,d}|^2}d\chi \nonumber \\
& = & e^{-\gamma/|G_{s,d}|^2}
\end{eqnarray}
and
\begin{eqnarray}
\tau_2 & = & \int_{\phi}^{\infty}\Bigg(\sum_{a=1}^{K_r}\frac{1}{|G_{a,d}|^2}e^{-\chi/|G_{a,d}|^2}q_a+\frac{1}{|G_{s,d}|^2}e^{-\chi/|G_{s,d}|^2}q_0\Bigg)d\chi \nonumber \\
& = & \sum_{a=1}^{K_r}q_a e^{-\phi/|G_{a,d}|^2}+q_0 e^{-\phi/|G_{s,d}|^2}.
\end{eqnarray}

As in (\ref{r-app-amc}), let $f$ be the dimensionless effective rate of the outer code for the concatenated coding strategy under consideration.  We see that the throughput $R_{sm}$ is approximated by
\begin{equation}\label{r-app-sm}
R_{app,sm} = \frac{f}{2}\tau_1+\frac{f}{4}(1-\tau_1)\tau_2.
\end{equation}

The computation of $q_a$ and $q_0$ is similar to that for $q_{a,i}$, and note that we can drop the subscript $i$, i.e. there is no dependence on distinct transmission modes here.  In particular, $q_a$ depends on $\rho_a$, which is the probability that relay $a$ decodes the source message, instead of $\rho_{a,i}$.  We have
\begin{eqnarray}
q_a & = & \int_{\gamma}^{\infty}\frac{1}{|G_{s,a}|^2}e^{-\chi/|G_{s,a}|^2}d\chi \nonumber \\
& = & e^{-\gamma/|G_{s,a}|^2}.
\end{eqnarray}

In Examples \ref{opt-fbk-prb} and \ref{opt-amc-swp} we will maximize the approximations in \eqref{r-app-amc} and \eqref{r-app-sm} to obtain optimal values of the contention probability and the switching point for adaptive modulation and coding.  Note that \eqref{r-app-amc} and \eqref{r-app-sm} assume that the receiver successfully decodes a packet after its received SNR exceeds a certain threshold.  In particular, SNR-threshold decoding yields neither a lower nor an upper bound on the throughput.  SNR-threshold decoding actually approximates the throughput, since successful decoding depends on the specific coded bits that are received in error.  The dependence of successful decoding on the code structure implies that for received SNR values $\gamma_1 < \gamma_2$, successful decoding may occur at $\gamma_1$ while near-successful decoding may occur at $\gamma_2$.

Thus, we can maximize the approximations in \eqref{r-app-amc} and \eqref{r-app-sm} and obtain good intuition for the optimal values of the contention probability and the AMC switching point, as seen in Examples \ref{opt-fbk-prb} and \ref{opt-amc-swp}.

\newtheorem{opt-fbk-prb}{Example}[section]
\begin{opt-fbk-prb}\label{opt-fbk-prb}
Optimization of Contention Probability
\end{opt-fbk-prb}

\begin{quote}
For the single-mode transmission strategy, consider the two-iteration decoding limit case as described above.  Now consider a simple scenario where we have $K_r = 2$ relays and $K = 1$ minislot.  By evaluating (\ref{r-app-sm}) we find that
\begin{equation}
R_{app,sm} = \frac{f}{2}e^{-\gamma/|G_{s,d}|^2}+\frac{f}{4}(1-e^{-\gamma/|G_{s,d}|^2})(e^{-\phi/|G_{1,d}|^2}q_1+e^{-\phi/|G_{2,d}|^2}q_2+e^{-\phi/|G_{s,d}|^2}q_0),
\end{equation}
\begin{equation}
q_1 = p_1e^{-\gamma/|G_{s,1}|^2}((1-p_2)e^{-\gamma/|G_{s,2}|^2}+1-e^{-\gamma/|G_{s,2}|^2})
\end{equation}
and
\begin{equation}
q_2 = p_2e^{-\gamma/|G_{s,2}|^2}((1-p_1)e^{-\gamma/|G_{s,1}|^2}+1-e^{-\gamma/|G_{s,1}|^2}).
\end{equation}

We place one relay at $(x_1,y_1) = (25,10)$ and the other relay at $(x_2,y_2) = (75,-10)$.  We use the Worldwide Interoperability for Microwave Access (WiMAX) signaling bandwidth, which is roughly 9 MHz \cite{WireMANWorkGrp}, and given a noise floor of -204dB/Hz this yields a noise value $N_0 = -134$dB.  Consider a case where the transmit power is 110dB above the noise floor of $N_0 = -134$dB.  Then we have $|G_{s,1}|^2 = 10^{(-134+110)/10}\cdot (9.89\cdot 10^{-5})\cdot (26.9)^{-3} = 2.02\cdot 10^{-11} = |G_{2,d}|^2$ and $|G_{s,2}|^2 = 10^{(-134+110)/10}\cdot (9.89\cdot 10^{-5})\cdot (75.7)^{-3} = 9.09\cdot 10^{-13} = |G_{1,d}|^2$.  Also, $|G_{s,d}|^2 = 10^{(-134+110)/10}\cdot (9.89\cdot 10^{-5})\cdot (100)^{-3} = 3.94\cdot 10^{-13}$.  Let $f = 1912/2050$, $\gamma \approx 13$dB and $\phi \approx -6$dB.

We maximize $R_{app,sm}$ with respect to $p_1$ and $p_2$.  The maximizing values are $p_{1,max} = 1$ and $p_{2,max} = 0$ and the maximum value of $R_{app,sm}$ is 0.25933.  Simulation results yield a throughput $R_{sm}$ of 0.23076, which shows that this approximation is good.  The maximizing values $p_{1,max}$ and $p_{2,max}$ reveal an interesting guideline for system designers.  In a two-relay network, if a single mode is used for transmission, the relay that is closer to the source than to the destination should always send a ACK message to the source if it has decoded the source message.  The other relay should never send any ACK messages to the source even if it has decoded the source message.

Intuitively, since the relay that is closer to the source has a better chance of decoding the source message than the relay that is closer to the destination, it will be able to assist the source more often than the relay that is closer to the destination.  The likelihood that both relays have decoded the source message is low, so the relay that is closer to the source should always assist the source if it has decoded the source message.  Thus, the relay that is closer to the destination should never interfere with the other relay.

Again we consider the simple case of $K_r = 2$ relays and $K = 1$ minislot.  Recall that
\begin{equation}
R_{app,amc} = \frac{f}{3}p_{1,1}+\frac{2f}{3}p_{2,1}+\frac{f}{3}(1-p_{2,1})p_{2,2}+\frac{f}{6}(1-p_{1,1})p_{1,2}+\frac{2f}{9}(1-p_{1,1})p_{2,2}+\frac{2f}{9}(1-p_{2,1})p_{1,2}
\end{equation}
and we find that
\begin{equation}
p_{1,1} = e^{-\alpha/|G_{s,d}|^2}-e^{-\gamma_{swp}/|G_{s,d}|^2},
\end{equation}
\begin{equation}
p_{2,1} = e^{-\beta/|G_{s,d}|^2},
\end{equation}
\begin{eqnarray}
p_{1,2} & = & (e^{-\phi/|G_{1,d}|^2}-e^{-\gamma_{swp}/|G_{1,d}|^2})q_{1,1} +(e^{-\phi/|G_{2,d}|^2}-e^{-\gamma_{swp}/|G_{2,d}|^2})q_{2,1}+\\
& & (e^{-\phi/|G_{s,d}|^2}-e^{-\gamma_{swp}/|G_{s,d}|^2})q_{0,1}\nonumber
\end{eqnarray}
and
\begin{equation}
p_{2,2} = e^{-\gamma_{swp}/|G_{1,d}|^2}q_{1,2}+e^{-\gamma_{swp}/|G_{2,d}|^2}q_{2,2}+e^{-\gamma_{swp}/|G_{s,d}|^2}q_{0,2}
\end{equation}
along with
\begin{equation}
q_{1,1} = p_1e^{-\alpha/|G_{s,1}|^2}((1-p_2)e^{-\alpha/|G_{s,2}|^2}+1-e^{-\alpha/|G_{s,2}|^2}),
\end{equation}
\begin{equation}
q_{2,1} = p_2e^{-\alpha/|G_{s,2}|^2}((1-p_1)e^{-\alpha/|G_{s,1}|^2}+1-e^{-\alpha/|G_{s,1}|^2}),
\end{equation}
\begin{equation}
q_{1,2} = p_1e^{-\beta/|G_{s,1}|^2}((1-p_2)e^{-\beta/|G_{s,2}|^2}+1-e^{-\beta/|G_{s,2}|^2})
\end{equation}
and
\begin{equation}
q_{2,2} = p_2e^{-\beta/|G_{s,2}|^2}((1-p_1)e^{-\beta/|G_{s,1}|^2}+1-e^{-\beta/|G_{s,1}|^2}).
\end{equation}

Again, we place one relay at $(x_1,y_1) = (25,10)$ and the other relay at $(x_2,y_2) = (75,-10)$.  Assume that the transmit power is 110dB above the noise floor of $N_0 = -134$dB, and so $|G_{s,1}|^2 = 10^{(-134+110)/10}\cdot (9.89\cdot 10^{-5})\cdot (26.9)^{-3} = 2.02\cdot 10^{-11} = |G_{2,d}|^2$ and $|G_{s,2}|^2 = 10^{(-134+110)/10}\cdot (9.89\cdot 10^{-5})\cdot (75.7)^{-3} = 9.09\cdot 10^{-13} = |G_{1,d}|^2$.  Also, $|G_{s,d}|^2 = 10^{(-134+110)/10}\cdot (9.89\cdot 10^{-5})\cdot (100)^{-3} = 3.94\cdot 10^{-13}$.  Let $f = 1912/2044$, $\gamma_{swp} \approx 4$dB, $\alpha \approx 3$dB, $\beta \approx 9$dB and $\phi \approx -6$dB.

We maximize $R_{app,amc}$ with respect to $p_1$ and $p_2$.  The maximizing values are $p_{1,max} = 0$ and $p_{2,max} = 1$ and the maximum value of $R_{app,amc}$ is 0.42882.  Simulation results yield a throughput $R_{amc}$ of 0.4225, which shows that this approximation is good.  Again, the maximizing values $p_{1,max}$ and $p_{2,max}$ reveal an interesting guideline for system designers.  In a two-relay network, if adaptive modulation is being used and the average received power at the destination is high, the relay that is closer to the destination than to the source should always send a ACK message to the source if it has decoded the source message.  The other relay should never send any ACK messages to the source even if it has decoded the source message.

Intuitively, since the received power at the destination is high, both relays have a good chance of decoding the source message.  Thus, the relay that is closer to the destination should always contend to forward the source message since it has a better chance of using Mode 2 in time slot $t_2$ than the other relay, which is a throughput-maximizing decision.
\end{quote}

\newtheorem{opt-amc-swp}[opt-fbk-prb]{Example}
\begin{opt-amc-swp}\label{opt-amc-swp}
Optimization of AMC Switching Point
\end{opt-amc-swp}

\begin{quote}
Consider another simple scenario where we have $K_r = 1$ relay and $K = 1$ minislot.  As we only have one relay, we set its contention probability $p_1 = 1$.  Recall that
\begin{equation}
R_{app,amc} = \frac{f}{3}p_{1,1}+\frac{2f}{3}p_{2,1}+\frac{f}{3}(1-p_{2,1})p_{2,2}+\frac{f}{6}(1-p_{1,1})p_{1,2}+\frac{2f}{9}(1-p_{1,1})p_{2,2}+\frac{2f}{9}(1-p_{2,1})p_{1,2}
\end{equation}
and we find that
\begin{equation}
p_{1,1} = e^{-\alpha/|G_{s,d}|^2}-e^{-\gamma_{swp}/|G_{s,d}|^2},
\end{equation}
\begin{equation}
p_{2,1} = e^{-\beta/|G_{s,d}|^2},
\end{equation}
\begin{equation}
p_{1,2} = (e^{-\phi/|G_{1,d}|^2}-e^{-\gamma_{swp}/|G_{1,d}|^2})q_{1,1}+(e^{-\phi/|G_{s,d}|^2}-e^{-\gamma_{swp}/|G_{s,d}|^2})q_{0,1}
\end{equation}
and
\begin{equation}
p_{2,2} = e^{-\gamma_{swp}/|G_{1,d}|^2}q_{1,2}+e^{-\gamma_{swp}/|G_{s,d}|^2}q_{0,2}
\end{equation}
along with
\begin{equation}
q_{1,1} = e^{-\alpha/|G_{s,1}|^2}
\end{equation}
and
\begin{equation}
q_{1,2} = e^{-\beta/|G_{s,1}|^2}.
\end{equation}

We place the relay at $(x_1,y_1) = (50,0)$.  The transmit power is 110dB above the noise floor of $N_0 = -134$dB, and so $|G_{s,1}|^2 = 10^{(-134+110)/10}\cdot (9.89\cdot 10^{-5})\cdot (50)^{-3} = 3.15\cdot 10^{-12} = |G_{1,d}|^2$.  Also, $|G_{s,d}|^2 = 10^{(-134+110)/10}\cdot (9.89\cdot 10^{-5})\cdot (100)^{-3} = 3.94\cdot 10^{-13}$.  Let $f = 1912/2044$, $\gamma_{swp} \approx 4$dB, $\alpha \approx 3$dB, $\beta \approx 9$dB and $\phi \approx -6$dB.

We maximize $R_{app,amc}$ with respect to $\gamma_{swp}$.  The maximizing value is $\gamma_{swp,max} = \alpha \approx 3$dB and the maximum value of $R_{app,amc}$ is 0.36752.  Again, the maximizing value $\gamma_{swp,max}$ reveals an interesting guideline for system designers.  In a single-relay network, the AMC switching point should be set equal to the minimum SNR that is required for any receiving node $i$ to be able to decode the source message if Mode 1 is used.  Thus, we maximize our usage of Mode 2 which is equivalent to maximizing the throughput.
\end{quote}

\section{Simulation Results}\label{sim-res}

\subsection{Overhead Analysis}
Before we present various simulation results for this paper, a discussion regarding the impact of the overhead signaling of our relay selection strategy on the yielded throughput is in order.  We refer to Fig. \ref{framing-structure} for this discussion.  The question here is: does the overhead signaling make a noticeable impact on throughput?  We will cite some figures from the IEEE 802.11a standard \cite{11aStan:99} in the following discussion.  Note that all transmissions are preceded by training symbols for channel estimation, frequency offset correction and timing synchronization along with information that indicates the modulation, code-rate and length of the transmission.  Also, a guard interval that is equal to the propagation delay of the network in Fig. \ref{system-model} occurs after each node transmission.  If the distance between the source and the destination is $d_{s,d} = 100m$, then the propagation delay is $(100m)/(3\cdot 10^8 m/s)\approx 0.3\mu s$.

Fig. \ref{framing-structure} shows that each time slot consists of four time intervals and we discuss each of them here.  The first interval consists of either an ACK or a NACK message from the destination indicating either successful or unsuccessful recovery of the source message.  This ACK or NACK message consists of a flag bit, so by employing one OFDM symbol for the ACK or NACK message, the duration of this time interval is $20\mu s + 4\mu s + 0.3\mu s= 24.3\mu s$.

The second interval consists of the relay contention period, which contains a set of $K$ minislots.  During each minislot, each decoding relay sends a ACK message which consists of its relay ID number to the source with a certain probability.  For $K_r = 20$ relays, each ID number will require $\lceil\log 20 \rceil = 5$ bits.  Thus, one OFDM symbol can be employed for the ACK message.  The duration of each minislot is $20\mu s + 4\mu s + 0.3\mu s = 24.3\mu s$.  If there are $K = 3$ minislots, the duration of this time interval is $3\cdot 24.3\mu s = 72.9\mu s$.

The third interval consists of a message from the source indicating which relay, if any, has been chosen for the next time slot.  This message consists of the ID number of the chosen node.  As for the second interval, one OFDM symbol can be employed for this message, so the duration of this time interval is $20\mu s + 4\mu s + 0.3\mu s = 24.3\mu s$.

The fourth interval consists of data transmission by either the source or by one of the chosen relays.  For the RCPC coding strategy described in Section \ref{sel-rcpc}, the transmitting node is using one of the codes in a particular RCPC code family.  For the AMC strategy described in Section \ref{chase-comb-tm}, the transmitting node is sending the source message using either of two transmission modes.  The data sent during this interval is accounted for by the throughput expressions in (\ref{hagenauer-throughput}) and (\ref{r-app-amc}).  For reference, the minimum number and maximum number of OFDM data symbols that can be transmitted in a frame are $\lceil (16+8\cdot 1+6)/216 \rceil = 1$ and $\lceil (16+8\cdot 4095+6)/24 \rceil = 1366$, respectively.

Since both the RCPC and AMC strategies allow for variable-length data frames, we consider a data packet that consists of 24 data bits per OFDM symbol with LENGTH parameter set to 2048.  Then, the number of OFDM symbols is $\lceil (16+8\cdot 2048+6)/24 \rceil = 684$.  Each OFDM symbol contains a guard interval of $0.8\mu s$ and the duration of the data portion of each symbol is $3.2\mu s$.  Then, the data requires $684\cdot 3.2\mu s = 0.00219s$ to transmit, and the total duration of the guard intervals is $684\cdot 0.8\mu s = 0.000547s$.  Also, $20\mu s$ is required for training before the data transmission occurs.

The ratio of the total overhead during the fourth interval to the data duration during the fourth interval is $(0.000547s+20\mu s+0.3\mu s)/0.00219s \approx 25.9\%$.  Also, the ratio of the total overhead from all four intervals for each time slot to the data duration during the fourth interval is $(0.000547s+20\mu s+0.3\mu s+24.3\mu s+72.9\mu s+24.3\mu s)/0.00219s \approx 31.4\%$.  Thus, we conclude that the overhead signaling that is inherent to our decentralized relay selection protocol does not have a significant impact on the throughput expressions in (\ref{hagenauer-throughput}) and (\ref{r-app-amc}) compared to the inherent overhead that occurs during the fourth interval.

\subsection{Performance Impact of Varying System Parameters}\label{perf-impact}
While a joint optimization of all of the key system parameters would maximize the throughput, this is fairly difficult.  Instead, in this sub-section we provide insights as to how each of the key system parameters individually affects the throughput.  In this sub-section we consider relay selection with RCPC signaling.

For simulation purposes, we employ the path loss model described in \cite{ZhaVal:PracRelaNetwGene:Jan:05}.  Let $\mathcal{E}_{x}$ be the energy in the transmitted signal $\textbf{x}(w)$.  Also, let $\lambda_c$ be the carrier wavelength, $d_0$ denote the reference distance, $d_{b,i}$ denote the distance between transmitting node $b$ and receiving node $i$, and $\mu$ be the path loss exponent.  Thus, the average received energy at node $i$ is
\begin{eqnarray}
\mathcal{E}_i & = & \mathbb{E}(|h_{b,i,a}|^2)\mathcal{E}_{x} \\
& = & (\lambda_c/4\pi d_0)^2(d_{b,i}/d_0)^{-\mu}\mathcal{E}_{x}.
\end{eqnarray}

We adopt similar simulation parameters as those in \cite{ZhaVal:PracRelaNetwGene:Jan:05}.  Here, we employ a carrier frequency $f_c$ = 2.4GHz, $d_0$ = 1m, $d_{s,d}$ = 100m and $\mu$ = 3, where $d_{s,d}$ is the distance between the source and the destination.  We then uniformly distribute $K_r = 20$ relays in the region between the source and the destination such that each relay $i$ is $d_{i,d} < d_{s,d}$ meters from the destination.  BPSK modulation is used for all packet transmissions, and all of the relays and the destination use maximum-likelihood (ML) decoding.  Again we use the WiMAX signaling bandwidth of roughly 9 MHz.

The codes of rates $\{4/5, 2/3, 4/7, 1/2, 1/3\}$ from the $M = 6$ RCPC family in \cite{Hag:RateCompPuncConv:Apr:88} are used.  Concatenated coding is used here, where the outer code is a (255, 239) Reed-Solomon code with symbols from $GF(2^8)$ and can correct at most 8 errors.  The mother code for the RCPC family is a rate-1/3 convolutional code with constraint length 7 and generator polynomial (145 171 133) in octal notation, which is employed in the EDGE standard \cite{GSMWorlEDGEPlt}.

For each packet, the source transmits some subset of its bits in the first time slot such that this subset forms a codeword from the rate-$4/5$ code.  If decoding at the destination is unsuccessful, the selected relay transmits additional parity bits such that the destination can attempt to decode a codeword from the rate-$2/3$ code.  If decoding at the destination is still unsuccessful, the relay selection and parity forwarding continues until the destination attempts to decode a codeword from the mother rate-$1/3$ code.  If this final decoding step is unsuccessful, the packet is declared to be lost, which adversely affects the throughput in (\ref{hagenauer-throughput}).

In this section and in Sections \ref{thr-compare} and \ref{thr-amc}, we define the average received SNR at the destination as follows.  Assume that the source uses a transmit energy of $\mathcal{E}_{t}(\gamma)$ during time slot $t_1$ that yields an average SNR $\gamma$ at the destination.  Then, all transmitting nodes will use a transmit energy of $\mathcal{E}_{t}(\gamma)$ during all subsequent time slots.

\begin{figure}[tb]
\begin{center}
\includegraphics[width=3.0in]{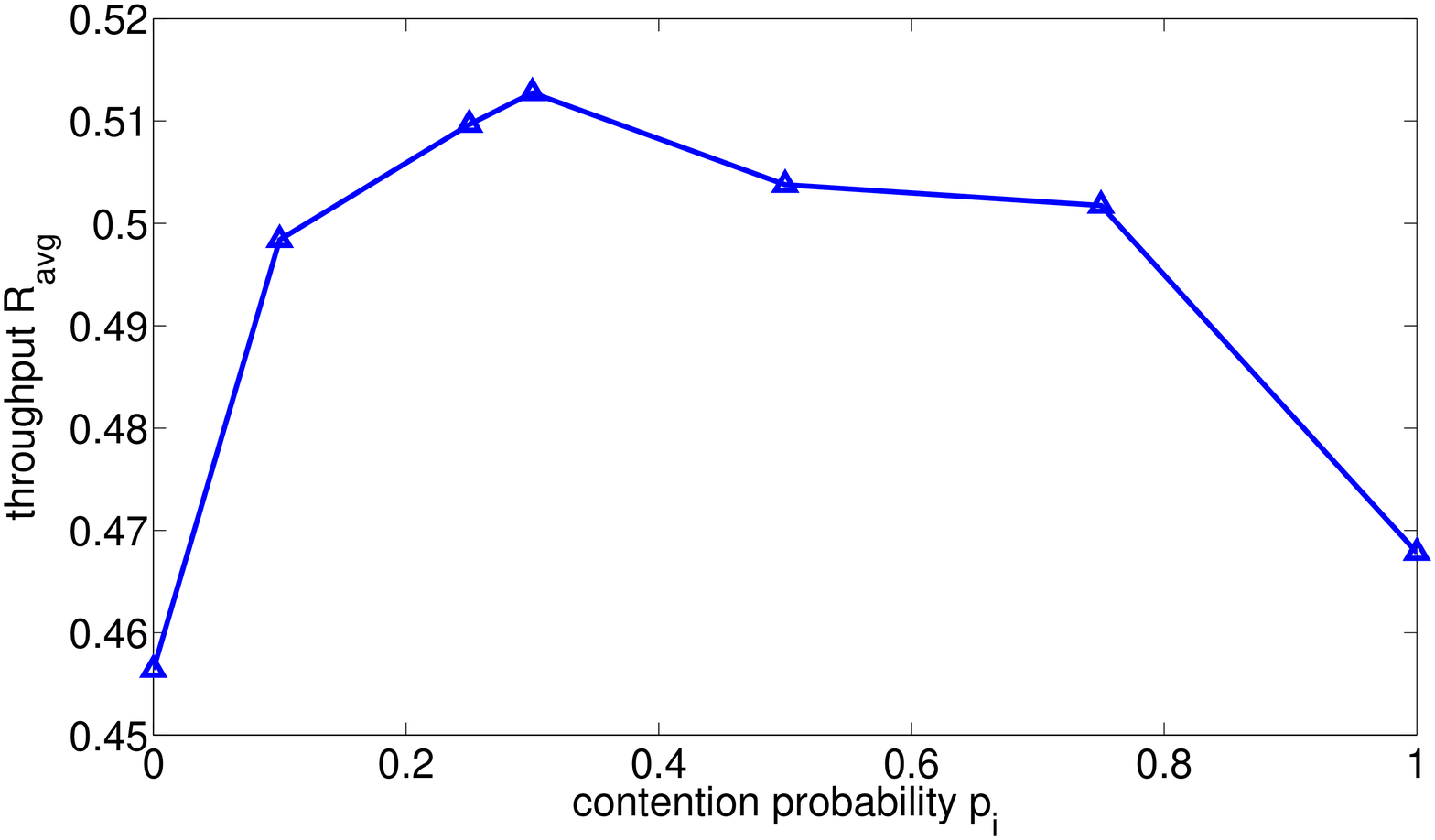}
\end{center}
\caption{Throughput as a function of contention probability for RCPC family with $M = 6$, rates $\{4/5,2/3,4/7,1/2,1/3\}$ and $d_{s,d} = 100m$.}
\label{fbk-prob}
\end{figure}

Fig. \ref{fbk-prob} shows how the throughput $R_{avg}$ yielded by the $\textit{ID}$ strategy varies with the contention probability $p_i$.  Here we fix $K = 10$ minislots and set the channel feedback threshold $\eta_{opp} = -91$dB.  The average received SNR at the destination is 2dB.  The throughput is maximized around $p_i = 0.3$.

The observed throughput performance has a nice intuitive explanation.  For large values of the contention probability $p_i$, each relay node $i \in \mathcal{R}_{sel}$ is more likely to send a ACK message to the source during each minislot $b$, which increases the likelihood of a collision during minislot $b$; this increases the likelihood that no relays will be selected during the entire contention period and that the source will end up forwarding the next set of parity bits to the destination.  For small values of the contention probability $p_i$, each relay node $i \in \mathcal{R}_{sel}$ is less likely to send a ACK message to the source during each minislot $b$, which decreases the likelihood of successful contention in minislot $b$ and increases the likelihood that the source will end up forwarding the next set of parity bits to the destination.

\begin{figure}[tb]
\begin{center}
\includegraphics[width=3.0in]{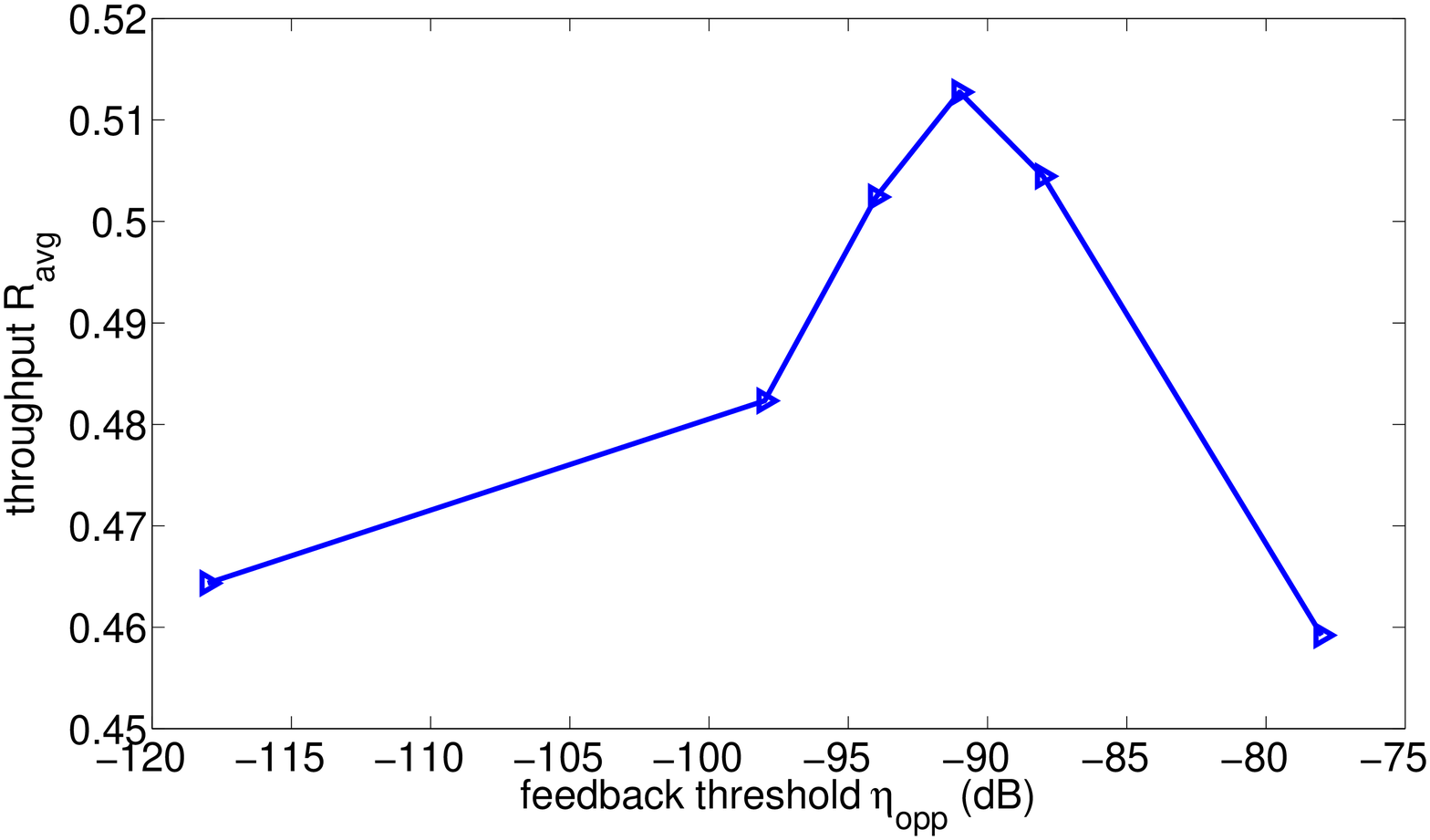}
\end{center}
\caption{Throughput as a function of feedback threshold for RCPC family with $M = 6$, rates $\{4/5,2/3,4/7,1/2,1/3\}$ and $d_{s,d} = 100m$.}
\label{eta-thresh}
\end{figure}

Fig. \ref{eta-thresh} shows how the throughput $R_{avg}$ yielded by the $\textit{ID}$ strategy varies with the channel feedback threshold $\eta_{opp}$.  Here we fix $K = 10$ minislots and set the contention probability $p_i = 0.1$.  The average received SNR at the destination is 2dB.  We see that the throughput is maximized around $\eta_{opp} = -91$dB.  The observed performance can be intuitively explained as follows.  For large values of the feedback threshold $\eta_{opp}$, $\|\mathcal{R}_{sel}\|$ is small, which decreases the likelihood of successful contention in minislot $b$.  For small values of the feedback threshold $\eta_{opp}$, $\|\mathcal{R}_{sel}\|$ is large, which increases the likelihood of a collision in minislot $b$.

\begin{figure}[tb]
\begin{center}
\includegraphics[width=3.0in]{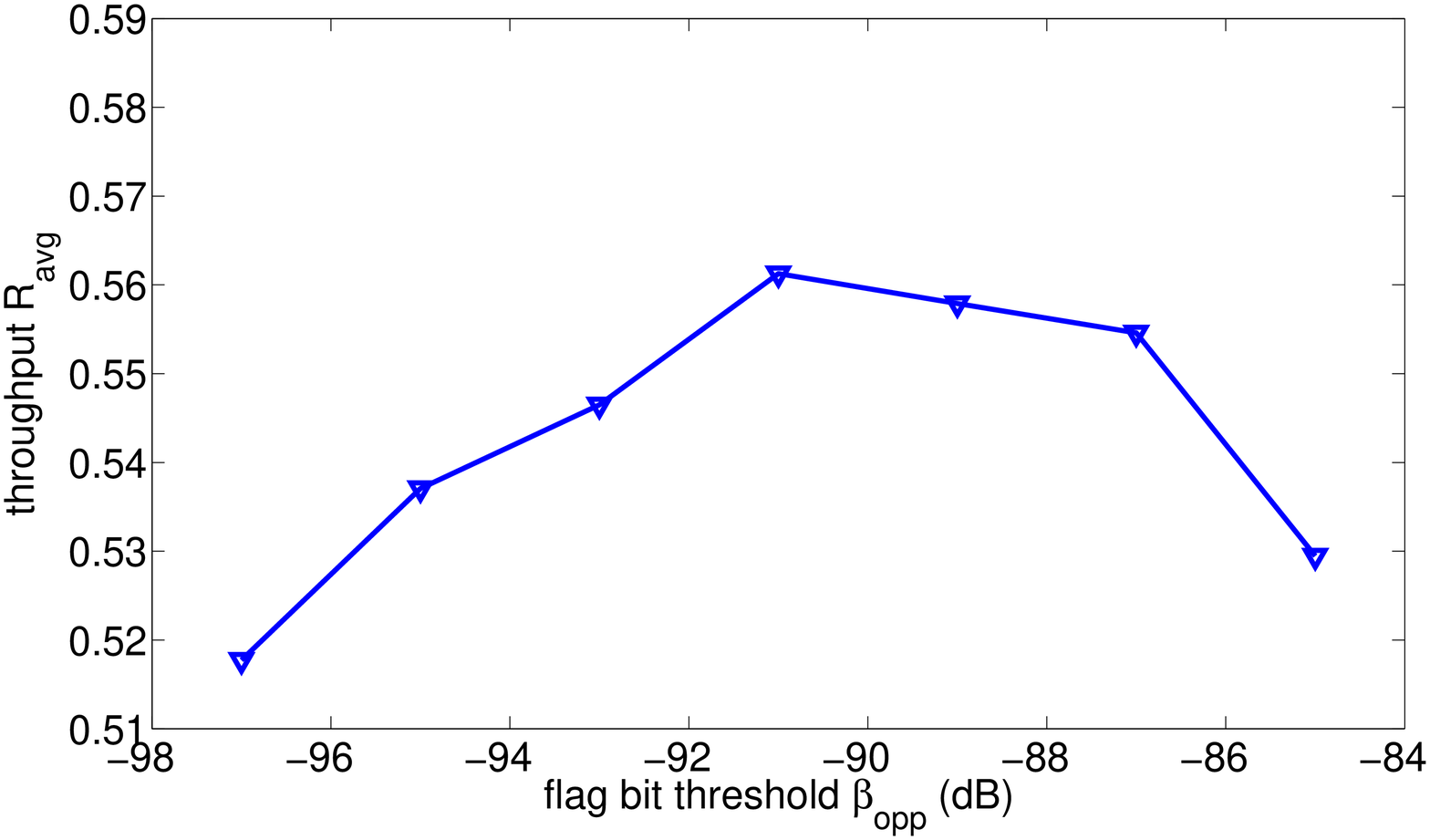}
\end{center}
\caption{Throughput as a function of flag bit threshold for average received SNR of 8dB, $K_r = 10$, $K = 3$ and $p_i = 0.3$.}
\label{beta-thresh}
\end{figure}

Fig. \ref{beta-thresh} illustrates the throughput of the $\textit{ID-CSI-1}$ strategy for various values of the flag bit threshold $\beta_{opp}$.  Here $K_r = 10$ relays and $K = 3$ minislots.  The average received SNR at the destination is 8dB.  We see that if $\beta_{opp}$ is close to $\eta_{opp}$, the performance of the $\textit{ID-CSI-1}$ strategy suffers since the $\textit{ID-CSI-1}$ strategy essentially reduces to the $\textit{ID}$ strategy.  Also, we see that if $\beta_{opp}$ is too large, the performance of the $\textit{ID-CSI-1}$ strategy suffers.  This is because the probability of selecting a decoding relay $i$ such that $|h_{i,d,a}|^2 > \beta_{opp}$ decreases as $\beta_{opp}$ increases, which causes the $\textit{ID-CSI-1}$ strategy to reduce to the $\textit{ID}$ strategy again.  Thus, it is apparent that there is an optimal value of $\beta_{opp}$ that maximizes the throughput of the $\textit{ID-CSI-1}$ strategy.

\begin{figure}[tb]
\begin{center}
\includegraphics[width=3.0in]{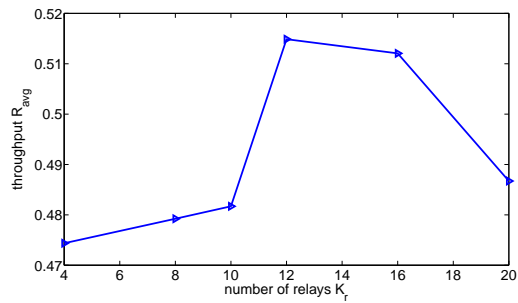}
\end{center}
\caption{Throughput as a function of number of relay nodes for average received SNR of 6dB, $K = 3$ minislots and $p_i = 0.3$.}
\label{relay-number}
\end{figure}

Fig. \ref{relay-number} illustrates the throughput of the $\textit{ID}$ strategy for a varying number of relay nodes.  We have $K = 3$ minislots and an average received SNR of 6dB at the destination.  We see that there is an optimal number of relay nodes for which the throughput is maximized.  Note that if the number of relay nodes is small, the probability that any of them decode the source message and send a ACK message to the source is also small.  On the other hand, if the number of relay nodes is large, the probability that at least two relays decode the source message and attempt to send a ACK message to the source in each minislot is also large, which increases the likelihood of a collision in each minislot.


\subsection{Throughput Comparison with HARBINGER Strategy}\label{thr-compare}
In this section we compare the throughput of the $\textit{ID}$ and $\textit{ID-CSI-1}$ strategies with the throughput of the HARBINGER approach in \cite{ZhaVal:PracRelaNetwGene:Jan:05}.  We also consider the throughput of a point-to-point transmission strategy where the source always forwards additional parity bits to the destination.  We set $\eta_{opp} = -91$dB, $p_i$ = 0.3, and $K = 10$ minislots, while the other simulation parameters are the same as in Section \ref{perf-impact}.

\begin{figure}[tb]
\begin{center}
\includegraphics[width=3.0in]{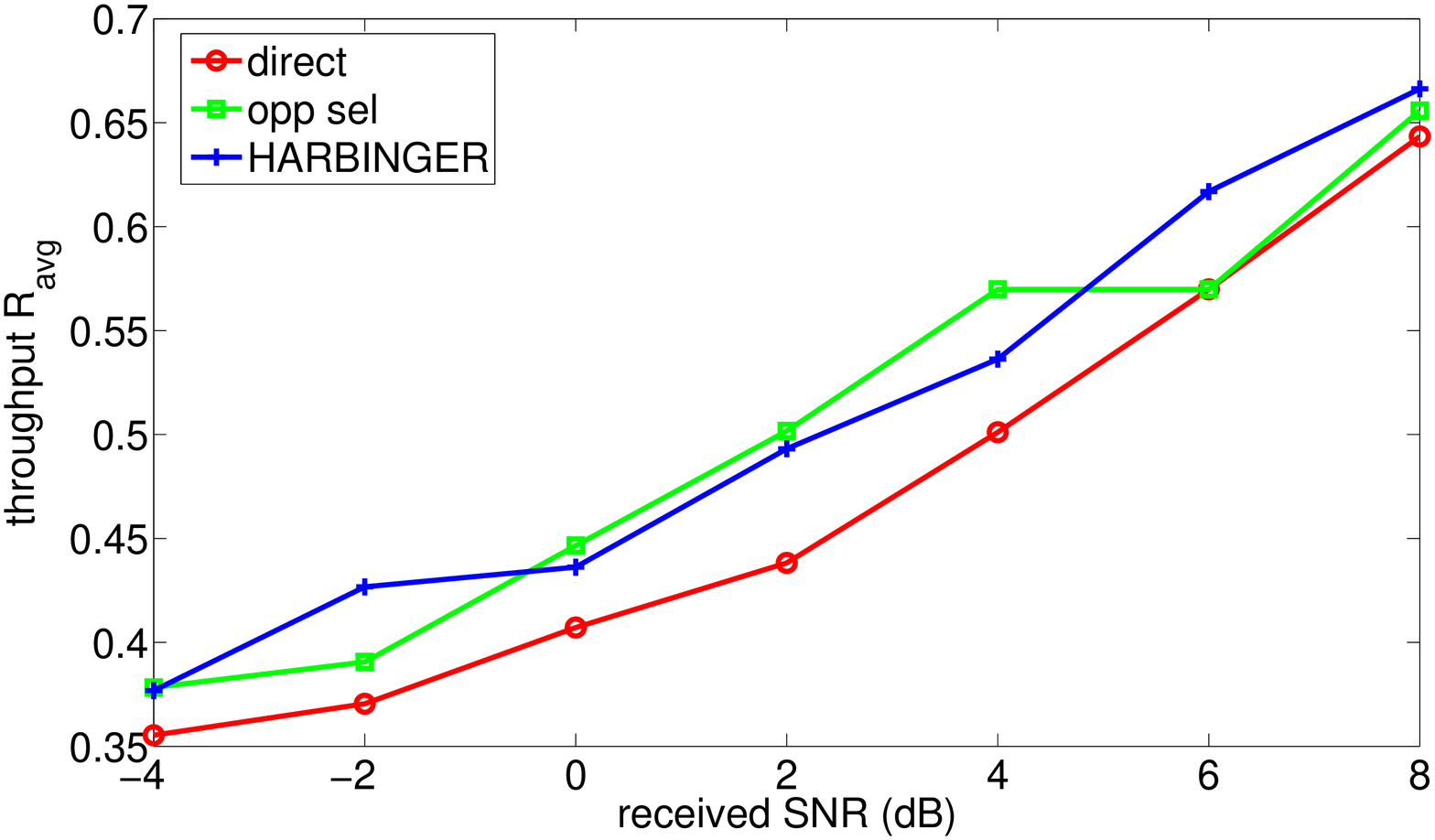}
\end{center}
\caption{Comparison with HARBINGER strategy in \cite{ZhaVal:PracRelaNetwGene:Jan:05} for RCPC family with $M = 6$, rates $\{4/5,2/3,4/7,1/2,1/3\}$ and $d_{s,d} = 100m$.}
\label{throughput}
\end{figure}

We see in Fig. \ref{throughput} that the $\textit{ID}$ strategy yields results that are comparable to those yielded by the HARBINGER approach, and in some cases, the decentralized strategy outperforms the HARBINGER approach.  This demonstrates that random access-based strategies can yield good performance.  Recall that the HARBINGER method optimizes the average received SNR at the destination by selecting the closest decoding relay to the destination to forward parity information.  This method, though, $\textit{does not necessarily select the}$ $\textit{decoding relay that would yield the highest instantaneous received SNR at the destination}$.  Thus, the decentralized strategy can outperform the HARBINGER method in some cases.

\begin{figure}[tb]
\begin{center}
\includegraphics[width=3.0in]{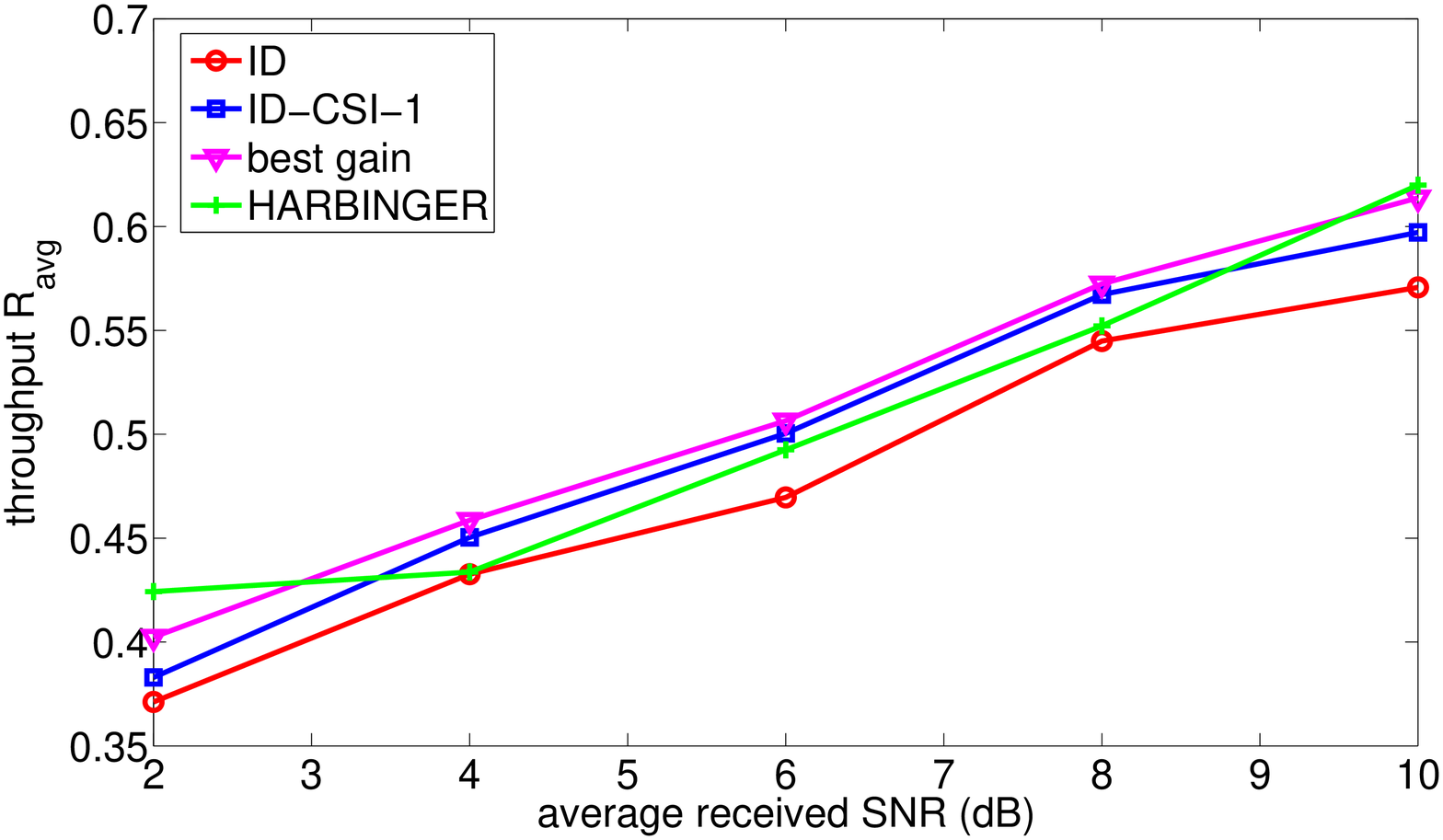}
\end{center}
\caption{Comparison of ID and ID-CSI-1 feedback strategies for RCPC family with $M = 6$, rates $\{4/5,2/3,4/7,1/2,1/3\}$ and $d_{s,d} = 100m$.}
\label{throughput2}
\end{figure}

Fig. \ref{throughput2} compares the throughput yielded by the $\textit{ID}$ and $\textit{ID-CSI-1}$ strategies.  We also plot the throughput yielded by the HARBINGER method and by a strategy that always selects the decoding relay with the best instantaneous channel gain to the destination to forward parity information.  We have $K = 10$ minislots.  For the $\textit{ID}$ and $\textit{ID-CSI-1}$ strategies, we set $\eta_{opp} = -91$dB and $\beta_{opp} = -86$dB.  We set the contention probability $p_i = 0.3$ for both strategies.  In addition, we set the winner selection probability $q = 0.75$ for the $\textit{ID-CSI-1}$ strategy.  We see that the $\textit{ID-CSI-1}$ strategy closes the performance gap between the $\textit{ID}$ strategy and the ``best-gain'' strategy.  Thus, using a limited amount of channel feedback improves the performance of our relay selection strategy.

\subsection{Throughput Performance of Adaptive Modulation Approach}\label{thr-amc}
We compare the throughput of the adaptive modulation and coding strategy from Section \ref{sel-amc} with the single-mode strategy from Section \ref{thr-app} and the RCPC strategy employing the channel feedback approach from Section \ref{chan-fbk-rel-sel}.  We use the modulation/code pairs from Section \ref{sel-amc} and Section \ref{thr-app}.  We adopt many of the simulation parameters and network topology from Section \ref{perf-impact} with some key exceptions.  In particular, we use $K = 10$ minislots, $K_r = 20$ relays and set the contention probability $p_i = 0.1$ for all relays $i$.

\begin{figure}[tb]
\begin{center}
\includegraphics[width=3.0in]{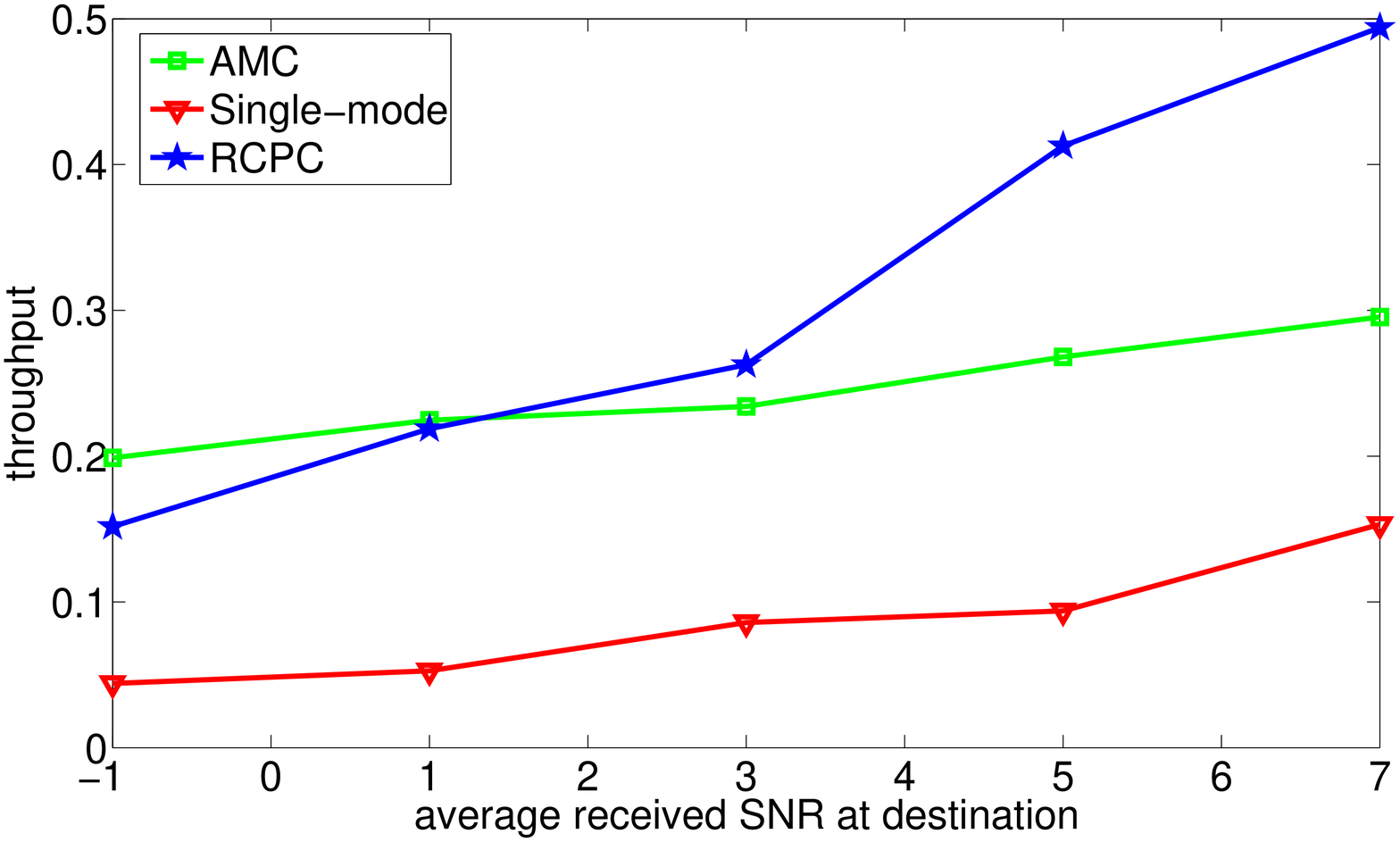}
\end{center}
\caption{Comparison of adaptive modulation, single-mode and RCPC strategies for $K = 10$ minislots, $K_r = 20$ relays and $p_i = 0.1$.}
\label{amc-sm-rcpc}
\end{figure}

In Fig. \ref{amc-sm-rcpc} we have a comparison of the throughput yielded by the adaptive modulation, single-mode and RCPC strategies.  Here, the SNR switching point is $\gamma_{swp} = 4$dB and we set a limit of 5 time slots before the destination stops trying to decode the source message.  We see that the adaptive modulation strategy significantly outperforms the single-mode strategy for this received SNR range.  For low received SNR values, the single-mode strategy suffers from high error rates because of the 16-QAM constellation that it uses.  On the other hand, the adaptive modulation strategy will use Mode 1 more often, and the combination of a rate-1/3 code and BPSK modulation will yield good error performance.  As the received SNR values increase, the single-mode strategy gradually performs better.  The throughput gap remains roughly constant, though, since the adaptive modulation strategy will use Mode 2 more often, and the use of a rate-2/3 code will lead to higher spectral efficiency.  Note that the RCPC strategy is outperformed by the adaptive modulation approach for low received SNR values, since the rate-1/3 mother code in the RCPC family is outperformed by diversity combining in this received SNR range.  On the other hand, for high received SNR values, the RCPC strategy outperforms the adaptive modulation approach since the rate-4/5 code in the RCPC family performs well in this received SNR range.

\section{Conclusion}
We have presented a decentralized relay selection protocol and described how channel feedback can be used to improve its performance in terms of throughput.  We have also derived an approximation of the throughput of our proposed adaptive modulation strategy and shown how maximization of this approximation can yield insights for system designers.  In addition, we have shown that our decentralized protocol yields throughput values that are comparable to those yielded by a centralized relay selection strategy that relies on location-based information \cite{ZhaVal:PracRelaNetwGene:Jan:05}.  By incorporating one bit of channel feedback in our selection strategy, we obtain throughput values comparable to those yielded by throughput-maximizing selection strategies that choose the decoding relay with the best channel gain to the destination.

Wireless network system design is a challenging problem, though, and the proposed selection protocol does not address many of the key issues that are inherent to it.  A more complete approach to performance optimization would involve a cross-layer strategy, where the physical-layer/MAC-layer approach in this paper is integrated with higher layers to yield improved performance.  For example, the buffer status of each relay could be evaluated as part of the selection process; if a relay has decoded the source message but is already handling a lot of traffic from other sources, it could decrease its contention probability.  Also, some aspects of the proposed approach could be optimized independently of other layers.  For example, the rate-1/3 mother code for the proposed adaptive modulation strategy could be chosen to be both systematic and have good minimum distance properties.

\end{document}